\documentstyle[pra,preprint,psfig,aps]{revtex}

\begin{document}
\draft
\title{Entanglement between external degrees of freedom of atoms via Bragg
deflection}
\author{Aeysha Khalique and Farhan Saif}
\address{Department of Electronics, Quaid-i-Azam university, 45320 Islamabad,\\
Pakistan.}
\maketitle

\begin{abstract}
We suggest that atoms undergoing Bragg deflection from a cavity field
introduce entanglement between their external degrees of freedom. The atoms
interact with an electromagnetic cavity field which is far detuned from
atomic transition frequency and is in superposition state. We provide a set
of experimental parameters in order to perform the suggested experiment
within the frame work of the presently available technology.
\end{abstract}


\newpage

\section{Introduction}

In 1935, Einstein, Podolsky and Rosen questioned the legitimacy of quantum
mechanics by using entangled states \cite{EPR}. Their work started enormous
philosophical discussions which led to improve the understanding of the
subject. The quantum entangled states have proved to be a foundation stone
in devising techniques to perform quantum computation \cite{Shor}, quantum
teleportation \cite{Telep} and quantum cryptography \cite{Cryp}. It has made
it vital to understand the roots of entanglement in quantum systems in its
details.

The generation of entanglement has been performed successfully between two
electromagnetic cavities \cite{David}, multimodes of a single
electromagnetic cavity \cite{Haroche}, internal states of atoms \cite{Atoms}%
, ions \cite{Ions} and Bose-Einstein condensates \cite{BEC}. Moreover
entanglement between angular momenta of a single atom \cite{Ang} and between
atoms in dark states \cite{Dark} has been suggested. In this paper we
present technique to develop entanglement between external degrees of
freedom of atoms which are defined by means of their momentum states. Our
suggested scheme relies on Bragg deflection of atoms from a cavity field. As
a manifestation of quantum duality, a matter wave passing through an optical
crystal at a certain angle displays Bragg deflection. We discuss that two
noninteracting matter waves incident on a cavity field in superposition,
generate entangled states comprising EPR-Bell basis by controlling
atom-field interaction times. We extend our suggested scheme to generate GHZ
\cite{GHZ,GHZ1} entangled states. Later, we discuss that presently available
experiments on atomic Bragg scattering from electromagnetic fields \cite
{Prichard,Rempe} may be considered to realize our theoretical work in
experiments.

We consider two supercooled atoms propagating with centre of mass momentum $%
P_{1}$ and $P_{2},$ and interacting simultaneously with a quantized standing
wave cavity field. The field inside the cavity is in a superposition state $%
1/\sqrt{2}(|0\rangle +|n_{0}\rangle ),$ where $|0\rangle $ is vacuum state
and $|n_{0}\rangle $ is any other Fock state$.$ We take the frequency of the
electromagnetic field far detuned from the transition frequency of the two
level atoms. Atom-field large detuning ensures that atoms do not exit the
cavity field in excited state and there is no spontaneous emission which
contributes photon in arbitrary direction. Thus interaction of atoms does
not alter the field state and both the atoms experience the field in the
same state during the time of interaction, that is either $|0\rangle $ or $%
|n_{0}\rangle $.

The effective Rabi frequency of each atom interacting with the field in
state $|n\rangle ,$ where $n$ is either $0$ or $n_{0},$ becomes $%
|g|^{2}n/2\triangle .$ Here, $g$ expresses coupling constant, and $\triangle
=\nu -w$ indicates detuning between the field frequency, $\nu ,$ and the
atomic transition frequency, $w.$ In order to study atomic deflection from
the cavity field we consider that the incident atoms propagate making an
angle $\theta $ with the normal to the cavity field. We apply Fresnel
approximation to the atomic motion, and, therefore we consider atomic
momentum component along the cavity field very small compared to the
component along the normal to the cavity field. Hence,
we treat the atomic motion along the normal to the cavity field classically.

We may express the evolution of the atoms interacting with the cavity field
by the effective Hamiltonian,
\begin{equation}
\hat{H}_{eff}=\frac{\hat{P}_{x_{1}}^{2}}{2M}+\frac{\hat{P}_{x_{2}}^{2}}{2M}-%
\frac{\hbar |g|^{2}}{2\triangle }\sum_{j=1,2}\stackrel{}{\hat{n}}\hat{\sigma}%
_{-}^{(j)}\hat{\sigma}_{+}^{(j)}\left( \cos 2k\hat{x}+1\right) ,
\label{Veff}
\end{equation}
which is obtained in presence of dipole approximation, rotating wave
approximation and secular approximation. Here, $\hat{P}_{x_{j}},$ is the
momentum operator describing momentum component along the cavity field of
each $j$th atom for $j=1,2,$ and $M$ indicates their respective mass.
Moreover, $\hat{\sigma}_{+}^{(j)}$ and $\hat{\sigma}_{-}^{(j)}$ are the
corresponding atomic raising and lowering operators and $\hat{n}$ describes
field number operator. Above Hamiltonian is separable for atoms 1 and 2.
This suggests that we may write the wave function of the system of two
atoms, $A_{1}$ and $A_{2}$ and field, $F$ as
\begin{equation}
|\Psi (A_{1},A_{2},F)\rangle =\frac{1}{\sqrt{2}}\sum_{l=-\infty }^{\infty
}\sum_{_{{}}l^{^{\prime }}=-\infty }^{\infty }\left[
C_{0,P_{l}}^{(1)}(t)C_{0,P_{l^{^{\prime
}}}}^{(2)}(t)|P_{l}^{(1)},P_{l^{^{\prime }}}^{(2)},0\rangle
+C_{n_{0},P_{l}}^{(1)}(t)C_{n_{0},P_{l^{^{\prime
}}}}^{(2)}(t)|P_{l}^{(1)},P_{l^{^{\prime }}}^{(2)},n_{0}\rangle \right] ,
\label{Psi}
\end{equation}
where, $C_{n,P_{l}}^{(1)}(t)$ $\left( C_{n,P_{l^{^{\prime
}}}}^{(2)}(t)\right) $ is the probability amplitude of atom 1 $\left(
2\right) $ exiting with momentum $P_{l}$ $\left( P_{l^{^{\prime }}}\right) $
in presence of field with $n$ photons. By comparing atomic scattering with
optical Bragg scattering \cite{Braggorg}, we \cite
{Marte,Mystre,Shore,Pichard,Gould} can develop a condition on initial
momentum of the incident atom, viz. $P_{l_{0}}=\frac{l_{0}}{2}\hbar k,$ for
which atomic Bragg scattering may occur. Here, $l_{0}=\pm 2,\pm 4$, $\pm 6$
etc., which correspond to first, second, third order of Bragg scattering,
respectively. By changing the atomic momentum component, $P_{l_{0}},$
parallel to the cavity field, we can change the order of Bragg scattering.
During interaction with the field for each complete Rabi cycle, momentum
transferred to the atom by the field is either $zero$ or $2\hbar k$ \cite
{Shore}$.$ Thus momentum of the exiting atom is given as $%
P_{l}=P_{l_{0}}+l\hbar k$, where, $l$ is an even integer.

Hence, by substituting the effective Hamiltonian, defined in Eq.~$\left( \ref
{Veff}\right) ,$ and the wave function given in Eq.~$\left( \ref{Psi}\right)
$ of our system in time dependent Schr\"{o}dinger equation$,$ we get
separate sets of infinite coupled rate equations for probability amplitudes $%
C_{n,P_{l}}^{\left( j\right) }$, for each $j$th atom$.$ We may express the
set of coupled rate equations as
\begin{equation}
i\frac{\partial C_{n,P_{l}}^{\left( j\right) }(t)}{\partial t}%
=w_{rec}l(l+l_{0})C_{n,P_{l}}^{\left( j\right) }(t)-\frac{\chi n}{2}\left(
C_{n,P_{l}+2\hbar k}^{\left( j\right) }(t)+C_{n,P_{l}-2\hbar k}^{\left(
j\right) }(t)\right) .  \label{cbf}
\end{equation}
Here, $w_{rec}=\frac{\hbar k^{2}}{2M}$ is recoil frequency of the atom and $%
\chi n=\frac{|g|^{2}n}{2\triangle }$ is effective Rabi frequency. In Bragg
deflection, recoil frequency of the deflected atom, is much larger than
effective Rabi frequency \cite{Marte,Arbab}, that is $w_{rec}\gg \chi n.$
Also, conservation of energy provides us $l=0\ $and\ $l=-l_{0},$ which leads
to two possible directions of propagation for the deflected atom, one with
momentum $P_{+l_{0}}^{\left( j\right) }$ and the other with momentum $%
P_{-l_{0}}^{\left( j\right) },$ respectively. We solve
the set of $l_{0}/2$ coupled equations, from $l=0$ to $l=-l_{0},$
adiabatically and obtain \cite{AySaif} two coupled equations as

\begin{eqnarray}
i\frac{\partial C_{n,P_{+l_{0}}}^{\left( j\right) }}{\partial t}
&=&A_{n}C_{n,P_{+l_{0}}}^{\left( j\right) }\left( t\right) -\frac{1}{2}%
B_{n}C_{n,P_{-l_{0}}}^{\left( j\right) }\left( t\right) ,  \label{r1} \\
i\frac{\partial C_{n,P_{-l_{0}}}^{\left( j\right) }}{\partial t}
&=&A_{n}C_{n,P_{-l_{0}}}^{\left( j\right) }\left( t\right) -\frac{1}{2}%
B_{n}C_{n,P_{+l_{0}}}^{\left( j\right) }\left( t\right) ,  \label{r2}
\end{eqnarray}
where,
\begin{equation}
A_{n}=\left\{
\begin{array}{r@{\quad \quad}l}
-\frac{\chi n/2}{w_{rec}(l_{0}-2)(2)} & \text{for }l_{0}\neq 2, \\
0 & \text{for }l_{0}=2,
\end{array}
\right.   \label{eq:vapp}
\end{equation}
and
\begin{equation}
|B_{n}|=\left\{
\begin{array}{r@{\quad \quad}l}
\frac{\left( \chi n\right) ^{^{\frac{l_{0}}{2}}}}{\left( 2w_{rec}\right) ^{%
\frac{l_{0}}{2}-1}\left[ (l_{0}-2)(l_{0}-4)......4.2\right] ^{2}} & \text{%
for }l_{0}\neq 2, \\
\chi n & \text{for }l_{0}=2.
\end{array}
\right.
\end{equation}
From Eqs.~$\left( \ref{r1}\right) $ and $\left( \ref{r2}\right) $, we obtain
the probability amplitudes of atoms exiting with momentum $P_{+l_{0}}$ and
that of exiting with momentum $P_{-l_{0}}$ as
\begin{eqnarray}
C_{n,P_{+l_{0}}}^{\left( j\right) }(t) &=&e^{-iA_{n}t/2}\left[
C_{n,P_{+l_{0}}}^{\left( j\right) }(0)\cos \left( \frac{1}{2}B_{n}t\right)
+C_{n,P_{-l_{0}}}^{\left( j\right) }(0)\sin \left( \frac{1}{2}B_{n}t\right)
\right] ,  \label{Bragg4} \\
C_{n,P_{-l_{0}}}^{\left( j\right) }(t) &=&e^{-iA_{n}t/2}\left[
C_{n,P_{-l_{0}}}^{\left( j\right) }(0)\cos \left( \frac{1}{2}B_{n}t\right)
+C_{n,P_{+l_{0}}}^{\left( j\right) }(0)\sin \left( \frac{1}{2}B_{n}t\right)
\right] .  \label{Bragg42}
\end{eqnarray}
Hence, as a result of Bragg deflection the probability of finding the
exiting atom in either of the two propagation directions flips, as a
function of interaction time $t,$ with frequency $|B_{n}|/2$ \cite{Logicgate}%
.

As defined in Eq.~$\left( \ref{Psi}\right) ,$ we may express the combined
state of the two deflected atoms and field, at any interaction time, $t,$ as
\begin{equation}
|\Psi (A_{1},A_{2},F)\rangle =\frac{1}{\sqrt{2}}\sum_{l=+l_{0},-l_{0}}%
\sum_{l^{^{\prime }}=+l_{0},-l_{0}}\left[
C_{0,P_{l}}^{(1)}(t)C_{0,P_{l^{^{\prime
}}}}^{(2)}(t)|P_{l}^{(1)},P_{l^{^{\prime }}}^{(2)},0\rangle
+C_{n_{0},P_{l}}^{(1)}(t)C_{n_{0},P_{l^{^{\prime
}}}}^{(2)}(t)|P_{l}^{(1)},P_{l^{^{\prime }}}^{(2)},n_{0}\rangle \right] ,
\end{equation}
where, $C_{n,P_{+l_{0}}}^{\left( j\right) }$ and $C_{n,P_{-l_{0}}}^{\left(
j\right) }$ are defined in Eqs.~$\left( \ref{Bragg4}\right) $ and $\left(
\ref{Bragg42}\right) $ for $j=1,2.$

In order to generate entanglement between the atomic external degrees of
freedom, we prepare one atom in the momentum state $|P_{+l_{0}}^{(1)}\rangle
$ and the other atom in the initial momentum state $|P_{-l_{0}}^{\left(
2\right) }\rangle ,$ at $t=0.$ Thus, for the atom 1 and atom 2 the
probability amplitudes, given in Eqs.~$\left( \ref{Bragg4}\right) $ and $%
\left( \ref{Bragg42}\right) ,$ reduce to
\begin{eqnarray}
C_{n,P_{+l_{0}}}^{\left( 1\right) }(t) &=&C_{n,P_{-l_{0}}}^{\left( 2\right)
}(t)=e^{-iA_{n}t/2}\cos \left( \frac{1}{2}B_{n}t\right) , \\
C_{n,P_{-l_{0}}}^{\left( 1\right) }(t) &=&C_{n,P_{+l_{0}}}^{\left( 2\right)
}(t)=e^{-iA_{n}t/2}\sin \left( \frac{1}{2}B_{n}t\right) .
\end{eqnarray}
We infer that atoms, initially in state $|P_{+l_{0}}^{(1)}\rangle $ and $%
|P_{-l_{0}}^{\left( 2\right) }\rangle ,$ after a time of interaction $t=s\pi
/B_{n},$ with the cavity field in state $|n\rangle $ $=|n_{0}\rangle ,$ are
deflected along $|P_{-l_{0}}^{\left( 1\right) }\rangle $ and $%
|P_{+l_{0}}^{(2)}\rangle ,$ respectively. Here $s$ is an odd integer. Since,
both $A_{n}$ and $|B_{n}|,$ disappear for $n=0,$ this implies that when
these atoms pass through the cavity field in vacuum state, they stay
undeflected. This yields $C_{n,P_{+l_{0}}}^{\left( 1\right)
}(t)=C_{n,P_{-l_{0}}}^{\left( 2\right) }(t)=0,$ and, $C_{n,P_{-l_{0}}}^{%
\left( 1\right) }(t)C_{n,P_{+l_{0}}}^{\left( 2\right) }(t)=e^{-i\varphi }.$
The phase $\varphi $ depends on the order of Bragg scattering $l_{0}$ and
the cavity field photon number $n_{0}$ and is given as $\varphi =s\pi
A_{n_{0}}/B_{n_{0}}.$ Hence, we get our system in state
\begin{equation}
|\Psi \left( A_{1}\text{,}A_{2}\text{,}F\right) \rangle =\frac{1}{\sqrt{2}}%
\left[ |P_{+l_{0}}^{\left( 1\right) },P_{-l_{0}}^{\left( 2\right) },0\rangle
+e^{-i\varphi }|P_{-l_{0}}^{\left( 1\right) },P_{+l_{0}}^{\left( 2\right)
},n_{0}\rangle \right] ,
\end{equation}
which is a three party entangled state, that is, between two atomic external
degrees of freedom and cavity field. We may get the same entangled state if
one atom interacts for a time $t=s\pi /B_{n_{0}}$ with the cavity field,
whereas the other atom with a time difference of $2r\pi /B_{n_{0}},$ where $%
r $ is an even integer. This makes $\varphi =\left( s+r\right) \pi
A_{n_{0}}/B_{n_{0}}.$ We may extract the entanglement of the external
degrees of freedom of the atoms by making measurement over the cavity field
state, which yields,
\begin{equation}
|\Psi \left( A_{1}\text{,}A_{2}\right) \rangle =\frac{1}{\sqrt{2}}\left[
|P_{+l_{0}}^{\left( 1\right) },P_{-l_{0}}^{\left( 2\right) }\rangle
+e^{-i\varphi }|P_{-l_{0}}^{\left( 1\right) },P_{+l_{0}}^{\left( 2\right)
}\rangle \right] .  \label{Bell1}
\end{equation}
In case, we keep the interaction time of one atom as $s\pi /B_{n_{0}},$
while let the other atom interact for an interaction time different by $%
2r^{^{\prime }}\pi /B_{n_{0}}$, where $r^{^{\prime }}$ is an odd integer, we
get the entangled state, as
\begin{equation}
|\Psi \left( A_{1}\text{,}A_{2}\right) \rangle =\frac{1}{\sqrt{2}}\left[
|P_{+l_{0}}^{\left( 1\right) },P_{-l_{0}}^{\left( 2\right) }\rangle
-e^{-i\varphi ^{^{\prime }}}|P_{-l_{0}}^{\left( 1\right)
},P_{+l_{0}}^{\left( 2\right) }\rangle \right] ,  \label{Bell2}
\end{equation}
where, $\varphi ^{^{\prime }}=\left( s+r^{^{\prime }}\right) \pi
A_{n_{0}}/B_{n_{0}}.$

We may generate the other two entangled states of the Bell basis by
preparing the two atoms in the same initial momentum states $%
|P_{+l_{0}}\rangle $ or $|P_{-l_{0}}\rangle $, and let them interact with
the cavity field. In case the field is in vacuum state, $|0\rangle ,$ the
incident atoms pass undeflected, whereas, in presence of field state, $%
|n_{0}\rangle ,$ the probability amplitudes of the incident atoms oscillate
as a function of interaction time. Now following our above discussion, the
interaction of one atom for a time $s\pi /B_{n}$ and the other for a time
difference of $2r\pi /B_{n},$ and, later, a measurement over the cavity
field state, leads us to the entanglement of the external degrees of freedom
of atoms as
\begin{equation}
|\Psi \left( A_{1}\text{,}A_{2}\right) \rangle =\frac{1}{\sqrt{2}}\left[
|P_{+l_{0}}^{\left( 1\right) },P_{+l_{0}}^{\left( 2\right) }\rangle
+e^{-i\varphi }|P_{-l_{0}}^{\left( 1\right) },P_{-l_{0}}^{\left( 2\right)
}\rangle \right] .  \label{Bell3}
\end{equation}
For the same initial conditions of the system but interaction times
different by an amount $2r^{^{\prime }}\pi /B_{n_{0}}$ of the two atoms with
the cavity field, we get the entangled state as
\begin{equation}
|\Psi \left( A_{1}\text{,}A_{2}\right) \rangle =\frac{1}{\sqrt{2}}\left[
|P_{+l_{0}}^{\left( 1\right) },P_{+l_{0}}^{\left( 2\right) }\rangle
-e^{-i\varphi ^{^{\prime }}}|P_{-l_{0}}^{\left( 1\right)
},P_{-l_{0}}^{\left( 2\right) }\rangle \right] .  \label{Bell4}
\end{equation}
Hence, our scheme leads us to generate complete set of Bell basis.

Equations~$\left( \ref{Bell3}\right) $ and $\left( \ref{Bell4}\right) $
provide a direct extension of our work to develop GHZ entangled state
between external degrees of freedom of atoms. We consider that initially we
prepare more than two atoms in the same momentum state $|P_{+l_{0}}\rangle $
or $|P_{-l_{0}}\rangle ,$ and, let them interact simultaneously with the
cavity field. Bragg deflection of the incident atoms ensures the generation
of GHZ state, $\frac{1}{\sqrt{2}}\left( |P_{+l_{0}}^{\left( 1\right)
},P_{+l_{0}}^{\left( 2\right) },...,P_{+l_{0}}^{\left( k\right) }\rangle \pm
e^{-i\varphi }|P_{-l_{0}}^{\left( 1\right) },P_{-l_{0}}^{\left( 2\right)
},...,P_{-l_{0}}^{\left( k\right) }\rangle \right) .$ Here an interaction
time of $s\pi /B_{n_{0}},$ leads to positive sign and $\varphi =ks\pi
A_{n_{0}}/2B_{n_{0}}$, however, any atom interacting for an interaction time
difference of $2r^{^{\prime }}\pi /B_{n_{0}}$ will lead to negative sign and
$\varphi =\left[ \left( k-1\right) s+2r^{^{\prime }}\right] \pi
A_{n_{0}}/2B_{n_{0}}$, where $k$ indicates the number of interacting atoms.

We may realize the suggested scheme in laboratory by using the experimental
set up of Ref. \cite{Rempe}. We propagate rubidium atoms of mass $%
M=1.42\times 10^{-25}$~Kg, through an optical quantum field of wavelength $%
\lambda =0.8~\mu m$. Therefore, the atoms experience a recoil frequency, $%
w_{rec}=2\pi \times 3.8$~kHz, while passing through the field, in presence
of a detuning by an amount $\triangle =2\pi \times 80$~MHz. We find that $%
g=2\pi \times 112$~kHz, such that, $\chi \approx 0.02w_{rec},$ which ensures
Bragg deflection of incident atoms. We make the times of interaction of the
two atoms with the cavity field different by controlling their initial
momentum components along the normal to the cavity field, as required to
generate entangled states expressed in Eqs.~$\left( \ref{Bell2}\right) $ and
$\left( \ref{Bell4}\right) .$ We may apply our suggested scheme in order to
engineer external degrees of freedom entanglement between different isotopes
of same material, between atoms of different materials and between an atom
and an ion.

\end{document}